%% file: main.tex
\documentclass[onecolumn,11pt,a4paper,reqno]{article}

\newif\ifdebug
\debugfalse
\usepackage{authblk}

\usepackage{graphicx,fullpage,xcolor}

\usepackage[utf8]{inputenc}        
\usepackage{lmodern}               
\usepackage[T1]{fontenc}           
\usepackage[english]{babel}

\usepackage{dsfont}

\usepackage{tabularx}

\usepackage{latexsym}              
\usepackage{amsmath}               
\usepackage{amssymb}               
\usepackage{amsfonts}              
\usepackage{amsthm}                

\usepackage{amsbsy}       
\allowdisplaybreaks
\usepackage{bbm}

\usepackage{tikz}                  
\usetikzlibrary{arrows,positioning,shapes}
\usepackage{tikzsymbols}

\usepackage{enumerate}             
\usepackage{enumitem}
\setlist{nolistsep}

\usepackage{xparse}                
\usepackage{hyperref}              
\usepackage{url}
\usepackage[capitalize,nameinlink]{cleveref}
\hypersetup{
    pdfpagemode=UseOutlines,
    linkbordercolor=[rgb]{1,1,1},
    filebordercolor=[rgb]{1,1,1},
    citebordercolor=[rgb]{1,1,1},
    menubordercolor=[rgb]{1,1,1},
    urlbordercolor=[rgb]{1,1,1},
    colorlinks=true,
    linkcolor=[rgb]{0,0.2,0.5},
    citecolor=[rgb]{0,0.35,0},
    urlcolor=[rgb]{0.6,0.2,0.4},
    breaklinks=true
}

\usepackage{color, xcolor}                 
\definecolor{darkgray}{rgb}{0.15,0.15,0.15}   
\definecolor{lightgray}{rgb}{0.94,0.94,0.94}  
\definecolor{lightlightgray}{rgb}{0.97,0.97,0.97}  
\definecolor{darkred}{rgb}{0.80,0.00,0.00}    
\definecolor{darkgreen}{rgb}{0.00,0.70,0.00}    
\definecolor{darkblue}{rgb}{0.00,0.00,0.70}    

\ifdebug
\usepackage[left]{showlabels}

\fi

\usepackage{graphicx}              
\graphicspath{{fig/}}              
\usepackage{caption}               
\usepackage{subcaption}            
\usepackage{multicol}              
\usepackage{float}                 
\usepackage{framed}                
\usepackage[framemethod=tikz]{mdframed}
\usepackage{mathtools}
\usepackage{enumitem}              
\usepackage{glossaries}            

\usepackage[bottom]{footmisc}

\numberwithin{equation}{section}
\theoremstyle{plain}  
\newtheorem{thm}{Theorem}[section]
\newtheorem{lem}[thm]{Lemma}
\newtheorem{conj}[thm]{Conjecture}
\newtheorem*{lem*}{Lemma}
\newtheorem{prop}[thm]{Proposition}
\newtheorem*{prop*}{Proposition}
\newtheorem{cor}[thm]{Corollary}
\newtheorem*{cor*}{Corollary}

\theoremstyle{definition}
\newtheorem{defn}[thm]{Definition}
\newtheorem*{defn*}{Definition}

\newtheorem*{notation*}{Notation}

\newtheorem*{problem*}{Problem}

\newtheorem*{question*}{Question}

\theoremstyle{remark}  

\newtheorem*{rmk*}{Remark}

\newtheorem*{ex*}{Example}

\theoremstyle{plain}
\newtheorem{innercustomgeneric}{\customgenericname}
\providecommand{\customgenericname}{}
\newcommand{\newcustomtheorem}[2]{%
  \newenvironment{#1}[1]
  {%
   \renewcommand\customgenericname{#2}%
   \renewcommand\theinnercustomgeneric{##1}%
   \innercustomgeneric
  }
  {\endinnercustomgeneric}
}

\newcustomtheorem{customthm}{Theorem}
\newcustomtheorem{customlem}{Lemma}
\newcustomtheorem{customcor}{Corollary}

\newcounter{protocol}

\crefname{protocol}{protocol}{protocols} 
\crefname{protocol}{Protocol}{Protocols}

\providecommand*{\hr}[1][class-arg]{%
    \hspace*{\fill}\hrulefill\hspace*{\fill}
    \vskip 0.65\baselineskip
}

\newcommand{\todo}[1]{\ifdebug \textcolor{red}{\small (#1) \marginpar{\small\textcolor{red}{to-do}}} \fi}

\newcommand{\norm}[1]{\left\lVert#1\right\rVert}
\newcommand{\brac}[1]{\left(#1\right)}

\newcommand{\cbrac}[1]{\left\{#1\right\}}

\newcommand{\abs}[1]{\left| #1 \right|}

\NewDocumentCommand{\prob}{mg}{
    \IfValueTF{#2}{
        P(#1 \,\vert\, #2)
    }{
        P(#1)
    }
}

\newcommand\restr[2]{{
  \left.\kern-\nulldelimiterspace 
  #1 
  \right|_{#2} 
  }}

\let\latexchi\chi
\makeatletter
\renewcommand\chi{\@ifnextchar_\sub@chi\latexchi}
\newcommand{\sub@chi}[2]{
  \@ifnextchar^{\subsup@chi{#2}}{\latexchi^{}_{#2}}%
}
\newcommand{\subsup@chi}[3]{
  \latexchi_{#1}^{#3}%
}
\makeatother

\newcommand{\R}{\mathbb{R}}

\newcommand{\F}{\mathbb{F}}

\newcommand{\bra}[1]{\langle#1|}
\newcommand{\ket}[1]{|#1\rangle}

\newcommand{\Tr}[1]{\mathrm{Tr}\left(#1\right)}

\makeatletter
\newcommand\RedeclareMathOperator{%
  \@ifstar{\def\rmo@s{m}\rmo@redeclare}{\def\rmo@s{o}\rmo@redeclare}%
}
\newcommand\rmo@redeclare[2]{%
  \begingroup \escapechar\m@ne\xdef\@gtempa{{\string#1}}\endgroup
  \expandafter\@ifundefined\@gtempa
     {\@latex@error{\noexpand#1undefined}\@ehc}%
     \relax
  \expandafter\rmo@declmathop\rmo@s{#1}{#2}}
\newcommand\rmo@declmathop[3]{%
  \DeclareRobustCommand{#2}{\qopname\newmcodes@#1{#3}}%
}
\@onlypreamble\RedeclareMathOperator
\makeatother

\RedeclareMathOperator{\Re}{Re}

\begin{document}

\input{preamble.tex}

\input{sections/Intro}
\input{sections/Prelims}
\input{sections/MatchingGame}

\input{sections/BipartiteMatching}

\input{sections/Knk}
\input{sections/PerfectMatching}

\bibliographystyle{alpha}

\bibliography{biblio}
\end{document}

%% file: preamble.tex
\title{Quantum Perfect Matchings}

\renewcommand\Affilfont{\itshape\small}

\author{David Cui}
\affil{Department of Mathematics, Massachusetts Institute of Technology}
\author{Laura Man\v{c}inska} 
\affil{Department of Mathematical Sciences, University of Copenhagen}
\author{Seyed Sajjad Nezhadi}
\affil{Joint Center for Quantum Information and Computer Science (QuICS), University of Maryland}
\author{David E. Roberson}
\affil{Department of Applied Mathematics and Computer Science,
Technical University of Denmark}

\date{}
\maketitle

\begin{abstract}
We investigate quantum and nonsignaling generalizations of perfect matchings in graphs using nonlocal games. Specifically, we introduce nonlocal games that test for $L$-perfect matchings in bipartite graphs, perfect matchings in general graphs and hypergraphs, and fractional perfect matchings. Our definitions come from the fact that these games are classical property tests for the corresponding matching conditions. We use the existence of perfect quantum and nonsignaling strategies for these games to define quantum and nonsignaling versions of perfect matchings. Finally, we provide characterizations of when graphs exhibit these extended properties:

\vspace{0.5em}

\begin{itemize}\setlength\itemsep{0.5em}
\item For nonsignaling matchings, we give a complete combinatorial characterizations. In particular, a graph has a nonsignaling perfect matching if and only if it admits a fractional perfect matching that has bounded value on triangles. 
\item In bipartite graphs, the nonsignaling $L$-perfect matching property is achieved exactly when the left component of the graph can be split into two disjoint subgraphs: one with a classical $L$-perfect matching and another with left-degree 2.
\item In the quantum setting, we show that complete graphs $K_n$ with odd $n \geq 7$ have quantum perfect matchings. We prove that a graph has a quantum perfect matching if and only if the quantum independence number of its line graph is maximal, extending a classical relationship between perfect matchings and line graph independence numbers. 
\item For bipartite graphs, we establish that the $L$-perfect matching game does not exhibit quantum pseudotelepathy, but we characterize the quantum advantage for complete bipartite graphs $K_{n,2}$. 
\item Additionally, we prove that deciding quantum perfect matchings in hypergraphs is undecidable and leave open the question of its complexity in graphs.
\end{itemize}

\end{abstract}

\tableofcontents

%% file: sections/Intro.tex
\section{Introduction} \label{Intro}


In this work, we investigate ``quantum'' notions of classical graph properties. To do this, we take the perspective of two-prover property testing, also known as nonlocal games. In this setup, there is a single verifier who wants to determine whether a particular object has a particular property. The verifier is allowed to quiz two spatially separated provers, often dubbed Alice and Bob, and cross-check their answers to verify whether the property holds. We say that a property test captures a particular property if the classical provers can convince the verifier with certainty if and only if the property holds. There is a vast literature on property testing \cite{Goldreich_2017}. In these tests, we can also allow for the provers to use quantum entangled strategies \cite{ito2012multiproverinteractiveproofnexp, 9719749, ji2020quantumsoundnessclassicallow, Natarajan_2017}. We can then say that the object has a quantum analog of the property if and only if the verifier can be perfectly convinced by the provers using a quantum entangled strategy. Previous works have established quantum analogs for several graph properties, such as quantum chromatic numbers, quantum independence numbers, and quantum graph homomorphisms and isomorphisms \cite{quantumHom, NSIso, quantumIso, QuantumColouring1, QuantumColouring2}. 

In some cases, such as $2$-colorability of graphs, the quantum and classical properties coincide. We refer to such tests as quantum sound, since quantum strategies cannot ``fool'' the classical test. However, in many interesting cases, the quantum property diverges from its classical counterpart, leading to new and well-motivated definitions of quantum properties. A particularly elegant example of this phenomenon is quantum isomorphism. While two non-isomorphic graphs may be quantum isomorphic, there is a striking combinatorial characterization of quantum isomorphism that naturally emerges as a relaxation of the classical test for graph isomorphism. In \cite{quantumIso}, the authors demonstrate that two graphs $G$ and $H$ are quantum isomorphic if and only if the homomorphism counts from any planar graph into them are identical, i.e., $\forall \,\text{planar } K , |\mathrm{Hom}(K,G)|= |\mathrm{Hom}(K,H)|$. This contrasts with the classical result, where two graphs $G$ and $H$ are isomorphic if and only if the homomorphism counts from any graph (not necessarily planar) into them are the same, i.e., $\forall K, |\mathrm{Hom}(K,G)|=|\mathrm{Hom}(K,H)|$. This combinatorial characterization is remarkable as the definition of a quantum isomorphism prima facie is not combinatorial in nature: the existence of a perfect quantum strategy is a highly continuous property in that it is defined through arbitrary high dimensional Hilbert spaces with arbitrary operators.

In this paper, we continue down this line of work and introduce natural nonlocal games to test for perfect matchings in graphs and hypergraphs, $L$-perfect matchings in bipartite graphs, and fractional perfect matchings. We then examine both quantum entangled strategies and nonsignaling strategies for these games. nonsignaling strategies generalize quantum strategies by imposing only one condition: Alice’s local marginal distributions must remain independent of Bob’s question, and vice-versa. Through these games, we define quantum and nonsignaling versions of perfect matchings, $L$-perfect matchings, and fractional perfect matchings. Additionally, we provide combinatorial characterizations for when a graph satisfies the nonsignaling properties, and we connect the quantum properties to other quantum graph properties, such as the quantum independence number.

\subsection{Our results}
In \cref{Games}, we define four separate synchronous nonlocal games that test for $L$-perfect matching for bipartite graphs, perfect matching for general graphs, fractional perfect matching and finally perfect matching for hypergraphs. We prove that these games characterize the classical graph properties. Namely, they have a perfect classical winning probability if and only if the underlying graphs have the requisite perfect matching property. Almost all other quantum graph properties in the literature are defined using the graph homomorphism game. Our tests represent a novel class of games that don't map onto the homomorphism games.

In \cref{Quantum}, we explore quantum and nonsignaling strategies for the $L$-perfect matching game. We prove that the game is quantum sound, by using Hall's theorem in graph theory and a result about the quantum chromatic number of clique graphs. As a consequence we also get that the fractional perfect matching test is quantum sound. We then go on to show that there are many bipartite graphs that only have perfect nonsignaling strategies for the game. We provide a full characterization for when a bipartite graph has this nonsignaling $L$-perfect matching property. In particular we prove the following theorem

\begin{thm}
    Let $G=(L\sqcup R,E)$ be a bipartite graph. Then the following are equivalent:
        \begin{enumerate}
            \item $\omega^{ns}(BPM_G) = 1$,
            \item $G^\#$ contains no lone vertices in $L^\#$,
            \item there exists a perfect matching subgraph $P \subset G$ and left-degree $2$ subgraph $S$ such that $P \sqcup S \subset G$ covering all of $L$.
        \end{enumerate}
\end{thm}

Where $G^\# = (L^{\#} \sqcup R^{\#}, E^{\#})$ is the graph where the degree $1$ vertices in $L$ and their neighbour in $R$ are iteratively removed until none remain.

Given that the $L$-perfect matching test is quantum sound a natural question, is whether there is any quantum advantage for these games at all. In \cref{Knk}, we explore the quantum values of the game for the bipartite complete graphs $K_{n,2}$. We fully characterize the quantum values by providing sum-of-square proofs of optimallity. We show that only $K_{3,2}$ has quantum advantage and that the optimal value is achieved via a synchronous strategy. In particular any synchronous strategy where the players three observables sum to $0$ are optimal for this game and those are the only ones. Interestingly for all $n \geq 4$ the optimally quantum and classical strategies for the games are non-synchronous. 

Finally, in \cref{NS} we turn to the perfect matching of general graphs. We show that, unlike fractional perfect matching and $L$-perfect matching, quantum strategies for the perfect matching game define a distinct property. In particular, $K_n$ for odd $n \geq 7$ have the quantum perfect matching property. We then provide a full characterization for the quantum perfect matching property. In particular we prove the following theorem

\begin{thm}
    Let $G$ be a graph. Then the following are equivalent:
    \begin{enumerate}
        \item $G$ has a quantum perfect matching.
        \item $L(G)$ has a projective packing of value $|V(G)|/2$.
        \item $\alpha_q(2L(G)) = |V(G)|$.
    \end{enumerate}
\end{thm}

This result mirrors the classical characterization of perfect matching. Where a graph $G$ has a perfect matching if and only if $\alpha(L(G)) = |V|/2$.

We then explore the nonsignaling perfect matching property. This further develops a distinct property from the quantum one. In particular, the cycle graphs $C_n$ for odd $n \geq 5$ have a nonsignaling perfect matching but not quantum or classical ones. We then provide a full characterization for the nonsignaling perfect matching property. In particular we prove the following theorem

\begin{thm}
    For a graph $G$, $\omega^{ns}(PM_{G}) = 1$ if and only if $G$ has a fractional perfect matching avoiding triangles.
\end{thm}

Finally, we show in the last part of the section that the quantum perfect matching property for hypergraphs is undecidable and in fact equivalent to that of deciding the quantum independence number of graphs.



\subsection{Further directions}

There are several further directions related to this work. 

\paragraph{Decidability of quantum perfect matching games and line graph problems}

Deciding whether a nonlocal game can be perfectly won with quantum strategies is generally undecidable \cite{Slofstra2019,mip*,pi2}. Additionally, it has been established that determining the quantum chromatic number, independence number, and quantum isomorphism are also undecidable \cite{ji2013binaryconstraintgameslocally, quantumIso,harris2023universalitygraphhomomorphismgames}. In \cref{und-hyper}, we demonstrate that quantum perfect matching for hypergraphs is undecidable. A natural open question is whether quantum perfect matching itself is undecidable. This turns out to be equivalent to whether the quantum independence number remain undecidable when restricted to only line graphs. 

One could more broadly ask: what about other graph properties restricted to line graphs, such as the chromatic number of line graphs? One natural approach is to use the classical reduction of Holyer which reduces 3SAT to edge coloring \cite{holyer}. Here, one 3SAT clause is mapped to a gadget graph where the edge coloring of the entire graph encodes the 3SAT clause assignment. Unfortunately, due to this type of gadget construction, the reduction does not ``quantize'' easily. To ensure that the reduction is quantum sound, i.e., if the 3SAT instance is not quantum satisfiable, then the reduced edge coloring instance is not quantum satisfiable, one would need to be able to simultaneously measure the entire gadget graph to recover an assignment to the 3SAT clauses. This requires commutativity. \cite{ji2013binaryconstraintgameslocally} gets around this issue by introducing a commutativity gadget which does not affect quantum satisfying solutions while enforcing that certain variables must commute. However, the approach of \cite{ji2013binaryconstraintgameslocally} was very specific to graph (vertex) coloring and such a commutativity gadget does not translate easily to edge problems. Hence, we ask whether one can construct such commutativity gadgets for the edge coloring problem. Furthermore, more generally, what kinds of constraint satisfaction problems admit commutativity gadgets? 

\paragraph{Characterization of the existence of quantum strategies for the perfect matching game} Man\v{c}inska and Roberson show that two graphs are quantum isomorphic exactly when their homomorphism counts from any planar graphs are equal \cite{quantumIso}. This is a completely combinatorial characterization of the existence of a perfect quantum strategy for the graph isomorphism game. In our work, we give a combinatorial characterization for the existence of a nonsignaling perfect strategy. Does there exist such a characterization for the perfect matching games?

\paragraph{Additional characterization for nonsignaling perfect matching} In \cref{thm:ns_characterization}, we show that the nonsignaling value of the perfect matching game on graph $G$ is $1$ if and only if $G$ has a fractional perfect matching avoiding triangles. A fractional perfect matching of $G$ is just a function $f : E(G) \to [0,1]$ such that around any vertex, the sum of $f$ is $1$. There is a classic theorem from graph theory that says that in fact we can restrict the codomain to just $\cbrac{0, 1/2, 1}$ which is equivalent to the fact that the graph can be decomposed into odd cycles and single matchings \cite{scheinerman13}. The proof of this theorem follows by choosing a fractional perfect matching with the smallest support and showing that this is the candidate fractional perfect matching with weights in $\cbrac{0, 1/2, 1}$. To accomplish this, certain subgraphs in the support are barred by showing that if they existed, then there is a transformation of the edge weights which would reduce the support even further. Unfortunately, not all of these transformations preserve the sum of edge weights on triangles and hence some adaptation is required for the case of fractional perfect matchings avoiding triangles. Nonetheless, we believe a statement like this should be true. In particular, 

\begin{conj}
    $G$ has a fractional perfect matching avoiding triangles if and only if $G$ has a fractional perfect matching avoiding triangles taking values in $\cbrac{0, 1/2, 1}$. 
\end{conj}

Note that this would imply that the graph $G$ also have a decomposition into odd cycles with size $\geq 5$ and single matchings.

\subsection*{Acknowledgements}
We would like to thank Connor Paddock and Benjamin Lovitz for helpful discussions. 

%% file: sections/Prelims.tex
\section{Preliminaries} \label{Prelims}

\subsection{Nonlocal games}

\begin{defn}
A \emph{nonlocal game} $\mathcal{G}$ is a tuple $(\mathcal{X}, \mathcal{A}, V)$ consisting of a finite set~$\mathcal{X}$ of inputs for Alice and Bob, and a finite set~$\mathcal{A}$ of outputs. As well as a verification function $V: \mathcal{X}\times \mathcal{X}\times \mathcal{A}\times \mathcal{A} \to \{0,1\}$.
\end{defn}

A nonlocal game is played by a verifier and two provers, Alice and Bob.
In the game, the verifier samples a pair $(x,y)$ uniformly at random and sends $x$ to Alice and $y$ to Bob.
Alice and Bob respond with $a$ and $b$, respectively.
They win if $V (x, y, a, b) = 1$.
The players are not allowed to communicate during the game, but they can agree on a strategy beforehand.
Their goal is to maximize their winning probability.

\begin{defn}
    A nonlocal game is called \emph{synchronous} if on simultaneously receiving the same questions Alice and Bob must respond identically to win, i.e. $V(x,x,a,b) = 0$ if $a \not = b$. Furthermore, we call a nonlocal game \emph{bisynchronous} if it is synchronous and additionally on receiving differing questions they may not respond with the same answer to win, i.e. $V(x,y,a,a) = 0$ if $x \not = y$.
\end{defn}

\begin{defn} \label{def:classical_strat}
    A \emph{classical strategy} $S$ for a nonlocal game $\mathcal{G}=(\mathcal{X}, \mathcal{A}, V)$ is a tuple $S=(f_A, f_B)$, consisting of maps $f_A : \mathcal{X} \to \mathcal{A}$ for Alice and $f_B : \mathcal{X} \to \mathcal{A}$ for Bob.
\end{defn}

\begin{defn} \label{def:tensor_strat} 
A \emph{quantum (tensor) strategy} $S$ for a nonlocal game $\mathcal{G}=(\mathcal{X}, \mathcal{A}, V)$ is a tuple $S=(\mathcal{H}_A,\mathcal{H}_B,\ket{\psi}, \{A_{xa}\}, \{B_{yb}\})$, consisting of finite dimensional Hilbert spaces~$\mathcal{H}_A$ and~$\mathcal{H}_B$, a bipartite state $\ket{\psi}\in \mathcal{H}_A\otimes \mathcal{H}_B$, PVMs $\{A_{xa}\}_{a\in \mathcal{A}}$ acting on $\mathcal{H}_A$ for each $x \in \mathcal{X}$ for Alice and PVMs $\{B_{yb}\}_{b\in \mathcal{A}}$ acting on $\mathcal{H}_B$ for each $y \in \mathcal{X}$ for Bob. Often we will drop the Hilbert spaces, and just write $S=(\ket{\psi}, \{A_{xa}\}, \{B_{yb}\})$.
\end{defn}

Here we restrict without loss of generality to pure states and projective measurements (PVMs).
For a strategy $S=(\ket{\psi}, \{A_{xa}\}, \{B_{yb}\})$, the probability of Alice and Bob answering $a, b$ when obtaining $x,y$ is given by $p(a,b|x,y)=\bra{\psi} A_{xa}\otimes B_{yb}\ket{\psi}$. Therefore, the \emph{winning probability} of a quantum strategy~$S$ for the nonlocal game $\mathcal{G}$ is given by
\begin{align*}
    \omega^*(S,\mathcal{G})=\sum_{x,y} \frac{1}{|\mathcal{X}^2|} \sum_{a,b}V(x,y,a,b)p(a,b|x,y)=\sum_{x,y} \frac{1}{|\mathcal{X}^2|} \sum_{a,b}V(x,y,a,b)\bra{\psi} A_{xa}\otimes B_{yb}\ket{\psi}.
\end{align*}
For a nonlocal game $\mathcal{G}$, we define the \emph{quantum value} $\omega^*(\mathcal{G})=\mathrm{sup}_{S}\omega^*(S,\mathcal{G})$ to be the supremum over all quantum tensor strategies compatible with $\mathcal{G}$. A game is said to exhibit pseudotelepathy if it has a quantum perfect strategy but no classical perfect strategy.

The tensor-product structure is a way of mathematically representing the locality of the players employing a quantum strategy in a nonlocal game. However, there is a more general way to model this nonlocality mathematically.

\begin{defn}\label{commuting_strat}
    A \emph{commuting operator strategy} $\mathcal{S}$ for a nonlocal game $\mathcal{G}=(\mathcal{X}, \mathcal{A}, V)$ is a tuple $\mathcal{S}=(\mathcal{H},\ket{\psi}, \{A_{xa}\}, \{B_{yb}\})$, consisting of a Hilbert space $\mathcal{H}$, a state $\ket{\psi}\in \mathcal{H}$, and two collections of mutually commuting PVMs $\{A_{xa}\}_{a\in \mathcal{A}}$ acting on $\mathcal{H}$ for each $x \in \mathcal{X}$ for Alice and PVMs $\{B_{yb}\}_{b\in \mathcal{A}}$ acting on $\mathcal{H}$ for each $y \in \mathcal{X}$ for Bob, i.e. $[A_{xa},B_{yb}]=0$ for all $a,b,x,y\in \mathcal{A}\times \mathcal{A}\times \mathcal{X}\times \mathcal{X}$. Like for quantum strategies, we will often omit the Hilbert space and write $\mathcal{S}=(\ket{\psi}, \{A_{xa}\}, \{B_{yb}\})$ for a commuting operator strategy.
\end{defn}

We can also define the commuting operator (also known as the quantum commuting) value of a nonlocal game $\omega^{qc}(\mathcal{G})=\sup_{S}\omega^{qc}(\mathcal{S},\mathcal{G})$ to be the supremum over all commuting operator strategies~$S$ compatible with $\mathcal{G}$. It is not hard to see that every quantum (tensor) strategy is a commuting operator strategy. The converse holds if we restrict our commuting operator strategies to be finite dimensional (i.e. $\mathcal{H}$ is finite dimensional). However, there are examples of nonlocal games $\mathcal{G}$ for which there is a perfect (wins with probability $1$) commuting operators strategy but no perfect tensor-product strategy, see for example \cite{Slofstra2019}.

\begin{defn} \label{def:ns_strat}
    A \emph{nonsignaling strategy} for a nonlocal game is any strategy where 
    $$\sum_a p(a,b,x,y) = \sum_a p(a,b,x',y) \text{ for every }b,y,x,x'$$ and similarly, $$\sum_b p(a,b,x,y) = \sum_b p(a,b,x,y') \text{ for every } a,x,y,y'.$$ 
\end{defn}

The nonsignaling value of a nonlocal game $\omega^{ns}(\mathcal{G})=\sup_{S}\omega^{ns}(\mathcal{S},\mathcal{G})$ to be the supremum over all nonsignaling strategies~$S$ compatible with $\mathcal{G}$. There are many games for which there is an optimal nonsignaling strategy but no quantum one. An example is the CHSH game for which any optimal quantum strategy wins with at most $\sim .85$ probability.

\begin{defn}
    We say that a strategy is \emph{synchronous} if $p(a,b,x,y) = 0$ whenever $x=y$ and $a\not = b$. Classical synchronous strategies are those for which Alice and Bob use the same map $f_A = f_B : \mathcal{X} \to \mathcal{A}$. Quantum synchronous strategies involve Alice and Bob using an identical set of measurements and a tracial state.
\end{defn}

We define the synchronous values of a nonlocal game $\omega^{t,s}(\mathcal{G})=\sup_{S}\omega^{t}(\mathcal{S},\mathcal{G})$ to be the supremum over all $t$-type synchronous strategies~$S$ compatible with $\mathcal{G}$, $t$ here being any of classical, quantum, commuting operator or nonsignaling. There are many games for which the synchronous and non-synchronous values differ \cite{synchval}. But for synchronous games we have the following result

\begin{thm}[\cite{synchval}] \label{thm:perf_synch_strat}
    If a synchronous game has a perfect strategy, then it also has a perfect synchronous strategy. This is true in all of the classical, quantum, commuting operator and nonsignaling settings.
\end{thm}

\subsection{Graph games}

\begin{defn}
    Given graphs $G = (V_G,E_G)$ and $H = (V_H,E_H)$, the homomorphism game $Hom(G,H)$ is the synchronous game with question set $V_G$ and answer set $V_H$. The players win if when they receive vertices in $G$ they respond with vertices in $H$ that preserve the adjacnecy relations between the players. I.e. on questions $g_A, g_B$ they respond with $h_A, h_B$ such that $g_A \sim g_B$ if and only if $h_A \sim h_B$.
\end{defn}

\begin{defn}
    The quantum chromatic number $\chi_q(G)$ of a graph $G$ is the smallest $c$ such that $Hom(G,K_c)$ has a perfect quantum strategy.

\end{defn}

The quantum chromatic number is sandwiched between the classical clique and chromatic numbers $\omega(G) \leq \chi_q(G) \leq \chi(G)$. \cite{QuantumColouring1}

\begin{defn}
    The quantum independence number $\alpha_q(G)$ of a graph $G$ is the largest $c$ such that $Hom(K_c, \bar{G})$ has a perfect quantum strategy, where $\bar{G}$ is the  graph complement of $G$.
\end{defn}

\begin{defn}
    The quantum clique number $\omega_q(G)$ of a graph $G$ is the largest $c$ such that $Hom(K_c, G)$ has a perfect quantum strategy.
\end{defn}

\begin{defn}
    Given graphs $G = (V_G,E_G)$ and $H = (V_H,E_H)$, the isomorphism game $Iso(G,H)$ is the synchronous game with question set $V_G \cup V_H$ and answer set $V_G \cup V_H$. The players must respond with a vertex in the opposite graph from that which they received a vertex. After which they will have $g_A, g_B \in G$ and $h_A, h_B \in H$. The players win if vertices in G (which may have been question or answer vertices for either) have the same adjacency relationship to those in $H$.
\end{defn}

\subsection{Graph theory}

\begin{defn}\label{defn:matching_def}
    Given a (hyper)graph $G = (V,E)$, a \emph{matching} $M \subset E$ is a set of pairwise non-adjacent edges. A \emph{perfect matching} is a matching which covers all vertices in $G$.
\end{defn}

\begin{defn}\label{defn:bipartite_matching_def}
    Given a bipartite graph $G = (L \sqcup R, E)$, an \emph{$L$-perfect matching} $M \subset E$ is a matching which covers all vertices in $L$.
\end{defn}

The following definition is a linear relaxation of the definition of a perfect matching.

\begin{defn}
    A graph $G=(V,E)$ has a \emph{fractional perfect matching} if there exists a function $f : E \to [0,1]$ such that for each vertex $v \in V$, $\sum_{(u,v) \in E} f \big( (u,v) \big) = 1$. 
\end{defn}

\begin{thm}[\cite{scheinerman13}]
    Given a graph with a fractional perfect matching, there is one where $f : E \to \{0,\frac{1}{2},1\}$.
\end{thm}

Given a graph $G = (V,E)$, denote $N_{G}(S) \subset V$ the set of vertices adjacent to the subset $S \subset V$.

\begin{thm}[Hall's marriage theorem] \label{thm:hall}
    Let $G = (L\sqcup R, E)$ be a bipartite graph. $G$ has a $L$-perfect matching if and only if every subset $S\subset L$ satisfies
    \[ \abs{S} \leq \abs{N_{G}(S)}. \]
\end{thm}

\begin{thm}[\cite{tutte}]
    A graph $G = (V,E)$ has a perfect matching if and only if for every subset $S \subset V$, the subgraph $G[V \setminus S]$ has at most $\abs{U}$ odd connected components.
\end{thm}

\begin{thm}[\cite{tutte}]
    A graph $G = (V,E)$ has a fractional perfect matching if and only if there is a collection of edges that form a disjoint covering of the vertices made up of matchings and odd cycles.
\end{thm}

Lastly, we define the notion of a line graph of a graph.

\begin{defn}
    Given a graph $G = (V,E)$, the \emph{line graph of $G$} denoted by $L(G)$ is the graph with vertices $E$ and edges $(e, f) \in E \times E$ such that $e \cap f \neq \emptyset$.
\end{defn}

\begin{prop}
    A graph $G = (V,E)$ has a perfect matching if and only if the independence number of its line graph is $\alpha(L(G)) = |V|/2$.
\end{prop}

%% file: sections/MatchingGame.tex
\section{Nonlocal games from perfect matchings} \label{Games}

In this section, we give nonlocal games corresponding to graph matching properties. These games are the main objects of study in this paper.

\subsection{The bipartite perfect matching game}

\begin{defn}
    Given a bipartite graph $G = (L \sqcup R, E)$, the \emph{bipartite $L$-perfect matching game $BPM_{G}$} is a synchronous game with question set $L$ and answer set $E$. The players win if and only if the question-answer pairs $(v_{1}, v_{2}) \in L \times L$ and $e_{1}, e_{2} \in E \times E$ satisfy
    \begin{enumerate}
        \item (adjacency) $v_{1} \in e_{1}$ and $v_{2} \in e_{2}$; and
        \item (consistency and edge disjointedness) either $e_{1} = e_{2}$ or $e_{1} \cap e_{2} = \emptyset$.
    \end{enumerate}
\end{defn}

The second condition says that if both players are given the same vertex then they have to give the same matching; and if they are given different vertices then they need to give different matchings. We can more eloquently write this condition as
\[ e_{1} \cap e_{2} \neq \emptyset \implies e_{1} = e_{2}. \]


We now show that these are indeed the natural games to define for the bipartite perfect matching property.

\begin{thm} \label{biclassical}
    Given a bipartite graph $G = (L \sqcup R, E)$, the game $BPM_{G}$ has a perfect classical strategy if and only if $G$ has an $L$-perfect matching. 
\end{thm}
\begin{proof}
    Let $G$ be a bipartite graph with an $L$-perfect matching $M$. On questions $(v_1,v_2)$, Alice and Bob will respond with $e_{1} \ni v_{1}$ and $e_{2} \ni v_{2}$, according to the matching $M$. Clearly, if $v_{1} = v_{2}$ then $e_{1} = e_{2}$. Now, since $M$ is a perfect matching if $v_{1} \neq v_{2}$ then $e_{1} \cap e_{2} = \emptyset$.
    
    Now suppose the game $BPM_G$ has a perfect classical strategy. Since $BPM_G$ is a synchronous game, by \cref{thm:perf_synch_strat} it must also have a perfect synchronous strategy where Alice and Bob both respond according to the same function $f : L \to E$. 
    Since $f$ is a perfect strategy and the defining conditions of $BPM_{G}$ are exactly those which define an $L$-perfect matching, $M = f(V)$ is an $L$-perfect matching. 
\end{proof}

\subsection{The perfect matching game}




We now give the natural extension of the bipartite perfect matching game to perfect matchings on an entire graph.

\begin{defn}
    Given a graph $G = (V, E)$, the \emph{perfect matching game $PM_{G}$} is a synchronous game with question set $V$ and answer set $E$. The players win if and only if the question-answer pairs $(v_{1}, v_{2}) \in V \times V$ and $e_{1}, e_{2} \in E \times E$ satisfy
    \begin{enumerate}
        \item (adjacency) $v_{1} \in e_{1}$ and $v_{2} \in e_{2}$; and
        \item (consistency and edge disjointedness) $e_{1} \cap e_{2} \neq \emptyset \implies e_{1} = e_{2}$.
    \end{enumerate}
\end{defn}

\begin{thm} \label{PMClassic}
Given a graph $G = (V, E)$ the game $PM_{G}$ has a perfect classical strategy if and only if $G$ has a perfect matching. 
\end{thm}
\begin{proof} 
The proof is identical to that of \cref{biclassical} except where $M$ is a perfect matching not an $L$-perfect matching.
\end{proof}

\subsection{The fractional perfect matching game}

In the graph theory literature, there is a notion of having a \emph{fractional perfect matching}. This is a relaxation of the usual notion of perfect matching where now the selected edges have a weight and the condition is that for all vertices, the weights of the edges incident to the vertex sum to $1$. We shall also define a nonlocal game which represents this property.

Given $G = (V,E)$, the \emph{bipartite double cover of $G$} is $G \times K_{2}$. We shall give a subset of the bipartite perfect matching games a different name. The motivation for this will be justified in the following theorems.

\begin{defn}
    Given a graph $G = (V, E)$, the \emph{fractional perfect matching game $FPM_{G}$} is the synchronous game $BPM_{G \times K_{2}}$.
\end{defn}

This definition is natural as we have the following lemma from graph theory.

\begin{lem}\label{lem:fpm_doublecover}
    $G$ has a fractional perfect matching if and only if $G \times K_{2}$ has a $L$-perfect matching.
\end{lem}
\begin{proof}
    Suppose $G$ has a fractional perfect matching. In particular then the graph has a covering made of disjoint odd cycles and matchings. To recover a $L$-perfect matching lift this cover onto the $G \times K_2$. Where the cycles are given some orientation and in the bipartite graph a vertex on the left is matched with a vertex on the right only if there was an outgoing edge in the original graph. Then this gives a matching since every left vertex is covered and its clearly disjoint.

    Now suppose $G \times K_2$ has a $L$-perfect matching. Then we can recover a a collection of disjoint odd cycles and matching on the original graph by mapping the edges in the $L$-perfect matching back onto $G$. Therefore, since there is a covering of $G$ by disjoint odd cycles and matching, $G$ has a perfect fractional matching.
\end{proof}

\begin{cor}
    $FPM_{G}$ has a perfect classical strategy if and only if $G$ has a fractional perfect matching.
\end{cor}
\begin{proof}
    By definition, $FPM_{G}$ has a perfect classical strategy if and only if $BPM_{G \times K_{2}}$ has a perfect classical strategy. Then by \cref{biclassical} and \cref{lem:fpm_doublecover}, $BPM_{G \times K_{2}}$ has a perfect classical strategy if and only if $G \times K_{2}$ has a perfect matching if and only if $G$ has a fractional perfect matching.
\end{proof}

We give another perspective on the fractional perfect matching games as a relaxation of $PM_G$. The consistency condition for the perfect matching game from before can be equivalently given by the following set of conditions:
If $e_{1} = (v_{1}, w_{1})$ and $e_{2} = (v_{2}, w_{2})$, then
\begin{enumerate}
    \item if $v_1 = v_2$ then $w_{1} = w_{2}$,
    \item if $v_1 \neq v_2$ then $w_{1} \neq w_{2}$, and
    \item $v_{1} = w_{2}$ if and only if $v_{2} = w_{1}$.
\end{enumerate}
Without the final ``symmetry'' condition this alternatively defines the fractional perfect matching game.

\subsection{The hypergraph perfect matching game}\label{hypergraph_def}

Continuing down this line of definitions, we can define the analogous perfect matching game for hypergraphs.

\begin{defn}
    Given a hypergraph $G = (V, E)$, the \emph{perfect matching game $PM_{G}$} is a synchronous game with question set $V$ and answer set $E$. Here $E$ represents a set of hyperedges. The players win if and only if the question-answer pairs $(v_{1}, v_{2}) \in V \times V$ and $e_{1}, e_{2} \in E \times E$ satisfy
    \begin{enumerate}
        \item (adjacency) $v_{1} \in e_{1}$ and $v_{2} \in e_{2}$; and
        \item (consistency and edge disjointedness) $e_{1} \cap e_{2} \neq \emptyset \implies e_{1} = e_{2}$.
    \end{enumerate}
\end{defn}

Again, we can show that classical perfect strategies for this game exactly correspond to hypergraph perfect matchings.

\begin{thm}
    Given a hypergraph $G = (V, E)$ the game $PM_{G}$ has a perfect classical strategy if and only if $G$ has a perfect matching. 
\end{thm}
\begin{proof}
    The proof is identical to that of \cref{PMClassic}.
\end{proof}






%% file: sections/BipartiteMatching.tex
\section{Quantum and nonsignaling bipartite perfect matching} \label{Quantum}

\subsection{Quantum bipartite perfect matchings}
Let $K_{n,k}$ be the complete bipartite graph. Then $BM_{K_{n,k}}$ is the same as the $k$-coloring game for the complete graph on $n$ vertices $K_n$. Therefore, for $k < n$ we have $\omega^*(BM_{K_{n,k}}) < 1$ since the quantum chromatic number of $K_n$ is $\chi_q(K_n) = n$ \cite{QuantumColouring1}. It turns out that no bipartite graph has quantum pseudo-telepathy for the bipartite perfect matching game precisely because of this fact.

\begin{thm}
    For any bipartite graph $G = (L\sqcup R, E)$, we have $\omega^*(BPM_{G}) = 1$ if and only if $\omega(BPM_{G}) = 1$.
\end{thm}
\begin{proof}
It suffices to prove that if $\omega^{\ast}\brac{BPM_{G}} = 1$ then $\omega\brac{BPM_{G}} = 1$. We do this by proving the converse. Suppose $\omega(BPM_G) < 1$ then from \cref{biclassical} $G$ has no $L$-perfect matching. Hence, by \cref{thm:hall}, there is some set $S \subset L$ such that $|N_G(S)| < |S|$. Let $n = |S|$ and $k = |N_G(S)|$. Considering the subgraph $G[S]$ induced by the subset $S$, we immediately see that
\[ \omega^{\ast}(BPM_{G[S]}) \leq \omega^{\ast}(BPM_{K_{n,k}}) < 1 \]
where the last inequality follows from the fact that $BPM_{K_{n,k}}$ is equal to the $k$-coloring game for $K_{n}$. Finally, since $\omega^{\ast}(BPM_{G[S]}) <1$ then $\omega^{\ast}(BPM_{G}) < 1$ since any perfect strategy for $BPM_{G}$ could just be restricted to $BPM_{G[S]}$.
\end{proof}

Since the fractional perfect matching game is just a subset of the bipartite perfect matching games, there is no pseudo-telepathy for all fractional perfect matching games as well.

\begin{cor}
    For any graph $G$, we have $\omega^*(FPM_{G}) = 1$ if and only if $\omega(FPM_{G}) = 1$.
\end{cor}

Even though quantum fractional and bipartite perfect matchings do not define new properties, we will see in \cref{Knk} these games can have quantum advantage. 

\subsection{Nonsignaling bipartite perfect matchings}
Let $G = (L\sqcup R,E)$ be a bipartite graph. Then let $G^\# = (L^{\#} \sqcup R^{\#}, E^{\#})$ be the graph where we iteratively removed the degree $1$ vertices in $L$ and their neighbours in $R$ until none remain. This process is well defined up to isomorphism. In this section we show that a bipartite graph $G$ has nonsignaling $L$-perfect matching if and only if $G^\#$ contains no lone vertices in $L^\#$.

In particular, the complete bipartite graphs $K_{n,k}$ with $k \geq 2$ satisfy the above property. Therefore, letting $n > k$, we get an infinite family of graphs with nonsignaling $L$-perfect matching but no classical $L$-perfect matching. 


Intuitively, this process just pairs off ``forced matchings'' and removes them from the graph. Once all of the ``forced matchings'' are removed, if there are any vertices in $G^{\#}$ which cannot be paired (i.e., they are isolated) then there can be no possible strategy. 

\begin{thm} \label{thm:ns-bipartite-characterization}
    Let $G=(L\sqcup R,E)$ be a bipartite graph. Then the following are equivalent:
        \begin{enumerate}
            \item $\omega^{ns}(BPM_G) = 1$,
            \item $G^\#$ contains no lone vertices in $L^\#$,
            \item there exists a perfect matching subgraph $P \subset G$ and left-degree $2$ subgraph $S$ such that $P \sqcup S \subset G$ covering all of $L$.
        \end{enumerate}
\end{thm}
\begin{proof}
    We shall first show that a left-degree 2 graph $G$ always has a nonsignaling perfect strategy. We shall explicitly define the nonsignaling correlation $p$. Take the correlation to be supported on $e_{1} \in E(v_{1})$ and $e_{2} \in E(v_{2})$, and set
    \[ p(e_{1}, e_{2}| v_{1}, v_{2}) = 
        \begin{cases}
            \frac{1}{2}, & \text{if } v_{1} = v_{2} \text{ and } e_{1} = e_{2} \\
            \frac{1}{2}, & \text{if } v_{1} \neq v_{2}, N(v_{1}) \cap N(v_{2}) \neq \emptyset, \text{and } e_{1} \cap e_{2} = \emptyset, N(v_{1}) \cap N(v_{2}) \subset e_{1} \cup e_{2} \\
            \frac{1}{4}, & \text{if } v_{1} \neq v_{2} \text{ and } N(v_{1}) \cap N(v_{2}) = \emptyset \\
            0, & \text{otherwise}.
        \end{cases}
    \]
    We see that this defines a probability distribution. For $v_{1} = v_{2}$,
    \[ \sum_{e_{1}, e_{2} \in E} p(e_{1}, e_{2}|v_{1}, v_{1}) = \sum_{ e \in E(v_{1}) } \frac{1}{2} = \frac{1}{2}\abs{E(v_{1})} = 1. \]
    For $v_{1} \neq v_{2}$, let $N(v_{1}) = \cbrac{w_{1}, w_{2}}$ and $N(v_{2}) = \cbrac{u_{1}, u_{2}}$. If, without loss of generality, $w_{1} = u_{1}$ and $w_{2} \neq u_{2}$ then 
    \[ \sum_{e_{1}, e_{2} \in E} p(e_{1}, e_{2}|v_{1}, v_{2}) = p((v_{1}, w_{1}), (v_{2}, u_{2})|v_{1}, v_{2}) + p((v_{1}, w_{2}), (v_{2}, u_{1})|v_{1}, v_{2}) = 1. \]
    If $N(v_{1}) = N(v_{2})$, then
    \[ \sum_{e_{1}, e_{2} \in E} p(e_{1}, e_{2}|v_{1}, v_{2}) = p( (v_{1}, w_{1}), (v_{2}, w_{2}) | v_{1}, v_{2} ) + p( (v_{1}, w_{2}), (v_{2}, w_{1}) | v_{1}, v_{2}) = 1. \]
    Finally, if $v_{1} \neq v_{2}$ and $N(v_{1} \cap N(v_{2}) = \emptyset$, then 
    \[ \sum_{e_{1}, e_{2} \in E} p(e_{1}, e_{2} | v_{1}, v_{2}) = \sum_{ e_{1} \in E(v_{1}), e_{2} \in E(v_{2}) } p(e_{1}, e_{2} | v_{1}, v_{2} ) = \frac{1}{4} \abs{E(v_{1})}\abs{E(v_{2})} = 1.  \]
    It is easy to see that this probability distribution satisfies perfectly all of the rules of the bipartite perfect matching game. Hence, we just need to check that it satisfies the nonsignaling condition. Indeed, for any $v_{1}, v_{2} \in L$ and $e_{2} \in E(v_{2})$ (or else $p$ is just $0$ and the nonsignaling condition trivially holds), we have 
    \[ \sum_{e_{1}} p(e_{1}, e_{2} | v_{1}, v_{2} ) = \frac{1}{2} \]
    in all cases. This shows that if one can find a perfect matching subgraph $P$ and left-degree $2$ subgraph $S$ such that $P$ and $S$ partition $L$, then we have a nonsignaling perfect strategy for $G$ and establishes (3) implies (1).

    Now, suppose that $G^{\#}$ contains a lone vertex in $L^{\#}$. A degree 1 vertex can only pick its unique edge in a perfect strategy, which also removes its neighboring right vertex from any other left vertices strategy. Therefore, the graph with the degree 1 left vertex and its neighbor removed has a perfect nonsignaling matching if and only if the original graph did. Then, it is clear that there is no nonsignaling strategy for $G$ since the lone vertex can not be matched in $G^\#$ and $G$ has a perfect nonsignaling matching if and only if $G$ does. This establishes (1) implies (2).

    Finally, we show that (2) implies (3). However, if $L^{\#}$ contains no lone vertices and the process has terminated then it must mean that $G^{\#}$ has left-degree $\geq 2$. Let $P$ be the removed degree $1$ edges and $S = G^{\#}$ and we obtain the desired condition for (3).
\end{proof}

From the above we also get that a graph $G=(V,E)$ has nonsignaling perfect fractional matching if $(G \times K_{2})^\#$ has no lone vertices on the left.

%% file: sections/Knk.tex
\section{Quantum bipartite matching $K_{n,2}$} \label{Knk}

In \cref{Quantum}, we showed that the bipartite matching game does not produce a separate quantum bipartite perfect matching property, this is fundamentally because $\omega^*(BPM_{K_{n,k}}) < 1$ for $k < n$.

However, this does not imply that quantum strategies do not exhibit advantage for the bipartite matching game. To study this we will look at the $K_{n,2}$ graphs and completely characterize their quantum and classical values. This is, in some sense, the most simple class of bipartite graphs one could consider which could have quantum advantage.

We first begin with deriving an expression for the winning probability in terms of quantum operators. We know that for a general strategy $p$, 
\begin{align*}
    \omega(BPM_{K_{n,2}}, p) &= \frac{1}{n^{2}} \sum_{v_{1},v_{2} \in [n]} \sum_{e_{1} \in E(v_{1}), e_{2} \in E(v_{2})} p(e_{1}, e_{2} | v_{1}, v_{2}) V_{BPM_{K_{n,2}}}(e_{1}, e_{2}, v_{1}, v_{2}) \\
                             &= \frac{1}{n^{2}} \sum_{v \in [n]} \sum_{e \in E(v)} p(e, e, | v, v) + \frac{1}{n^{2}} \sum_{v_{1} \neq v_{2} \in [n]} \sum_{\substack{ e_{1} \in E(v_{1}), e_{2} \in E(v_{2}) :\\ e_{1} \cap e_{2} = \emptyset} } p(e_{1}, e_{2}|v_{1}, v_{2}) \\
                             &= \frac{1}{n^{2}} \sum_{v \in [v]} \sum_{a \in [2]} p((v, a), (v, a) | v, v) + \frac{1}{n^{2}} \sum_{v_{1} \neq v_{2} \in [n]} \sum_{ a_{1} \neq a_{2} \in [2] } p((v_{1},a_{1}), (v_{2}, a_{2})|v_{1}, v_{2}).
\end{align*}
Note that once the question, which is a left vertex, is fixed, a right vertex completely determines the edge. Hence, for a quantum strategy, we have operators $\cbrac{ A_{va}}$ and $\cbrac{B_{ub}}$ with state $\rho$ such that
\[ \omega(BPM_{K_{n,2}}, p) = \frac{1}{n^{2}} \Tr{ \brac{  \sum_{v \in [v]} \sum_{a \in [2]} A_{va}B_{va} + \sum_{v_{1} \neq v_{2} \in [n]} \sum_{ a_{1} \neq a_{2} \in [2] } A_{v_{1}a_{1}}B_{v_{2}a_{2}} }\rho }. \]
Noting that
\begin{align*}
    2\omega(BPM_{K_{n,2}}, p) - 1 = &\frac{1}{n^{2}} \mathrm{Tr}\left( \left(  \sum_{v \in [v]} \brac{ \sum_{a \in [2]} A_{va}B_{va} - \sum_{a_{1} \neq a_{2} \in [2]} A_{va_{1}}B_{va_{2}} } \right. \right. \\ 
    & \qquad\qquad\qquad + \left. \left. \sum_{v_{1} \neq v_{2} \in [n]} \brac{ \sum_{ a_{1} \neq a_{2} \in [2] } A_{v_{1}a_{1}}B_{v_{2}a_{2}} - \sum_{a \in [2]} A_{v_{1}a}B_{v_{2}a} } \right)\rho \right).
\end{align*}
Defining $A_{v} := A_{v1} - A_{v2}$ and $B_{v} := B_{v1} - B_{v2}$, we see that
\begin{align*}
    \omega(BPM_{K_{n,2}}, p) &= \frac{1}{2n^{2}} \Tr{ \brac{ \sum_{v \in [n]} A_{v}B_{v} - \sum_{v_{1} \neq v_{2} \in [n]} A_{v_{1}}B_{v_{2}} } \rho } + \frac{1}{2} \\
                             &= \frac{1}{2n^{2}} \Tr{ \brac{ n^{2}I - \sum_{v_{1} \in [n]} A_{v_{1}}\brac{ \sum_{v_{2} \in [n]} (-1)^{\delta_{v_{1} = v_{2}}} B_{v_{2}} } } \rho }
\end{align*}

Lets start by looking at quantum synchronous strategies, we have that the winning probability is described by a single set of observables $\cbrac{A_{v}}$. Thus,
\begin{align*} 
    \omega(BPM_{K_{n,2}}, p) &= \frac{1}{2n^{2}} \Tr{ \sum_{v \in [n]} A_{v}^{2} - \sum_{v_{1} \neq v_{2} \in [n]} A_{v_{1}}A_{v_{2}} } + \frac{1}{2} \\
                             &= \frac{1}{2n^{2}} \Tr{ nI - \sum_{v_{1} \neq v_{2} \in [n]} A_{v_{1}}A_{v_{2}} } + \frac{1}{2} \\
                             &= \frac{1}{2n^{2}} \Tr{ 2nI - \sum_{v_{1}, v_{2} \in [n]} A_{v_{1}}A_{v_{2}} } + \frac{1}{2} \\
                             &= \frac{1}{2} + \frac{1}{n} - \frac{1}{2n^{2}} \Tr{ \brac{\sum_{v} A_{v} }^{2} }.
\end{align*}
This shows that whenever we have observables such that $\sum_{v} A_{v} = 0$, then we have a quantum synchronous strategy achieving value $\frac{1}{2} + \frac{1}{n}$. Hence we've shown the following lowerbound.

\begin{lem}
    $\omega^{\ast}(BPM_{K_{n,2}}) \geq \frac{1}{2} + \frac{1}{n}$ for all $n \geq 2$.
\end{lem}

Additionally, we have that
\[ \frac{1}{2} + \frac{1}{n} - \brac{ \frac{1}{2n^{2}} \sum_{v \in [n]} A_{v}^{2} - \sum_{v_{1} \neq v_{2} \in [n]} A_{v_{1}}A_{v_{2}} + \frac{1}{2} } = \frac{1}{2n^{2}} \brac{\sum_{v} A_{v}}^{2} \]
at the algebra level. 
This sum-of-squares decomposition proves $\frac{1}{2} + \frac{1}{n}$ is optimal for quantum synchronous strategies and therefore $\omega^{\ast,s}(BPM_{K_{n,2}}) = \frac{1}{2} + \frac{1}{n}$. 

Now we will upperbound the general quantum value of the $K_{n,2}$ games.


\begin{lem}
    We have the following upperbounds
    \begin{enumerate}
        \item $\omega^{\ast}(BPM_{K_{3, 2}}) \leq \frac{5}{6}$.
        \item $\omega^{\ast}(BPM_{K_{n, 2}}) \leq 1 - \frac{1}{n}$ for $n \geq 4$.
    \end{enumerate}
\end{lem}
\begin{proof}
    Consider the Frobenius norm with respect to some state $\rho$. We have that
    \begin{align*}
        \frac{1}{2n^{2}} \left\lVert n^{2} - \sum_{v_{1}} A_{v_{1}} \left( \sum_{v_{2}} (-1)^{\delta_{v_{1}=v_{2}}} B_{v_{2}} \right)  \right\rVert_{\rho} &\leq \frac{1}{2} + \frac{1}{2n^{2}} \left\lVert \sum_{v_{1}} A_{v_{1}} \left( \sum_{v_{2}} (-1)^{\delta_{v_{1}=v_{2}}} B_{v_{2}} \right)  \right\rVert_{\rho} \\
            &\leq \frac{1}{2} + \frac{1}{2n^{2}} \sum_{v_{1}}\norm{A_{v_{1}}}_{\rho}\left\lVert \left( \sum_{v_{2}} (-1)^{\delta_{v_{1}=v_{2}}} B_{v_{2}} \right)  \right\rVert_{\rho} \\
            &\leq \frac{1}{2} + \frac{1}{2n^{2}} \sum_{v_{1}}\left\lVert \left( \sum_{v_{2}} (-1)^{\delta_{v_{1}=v_{2}}} B_{v_{2}} \right)  \right\rVert_{\rho} \\
            &= \frac{1}{2} + \frac{1}{2n^{2}}\sqrt{n}\sqrt{ \sum_{v_{1}} \left\lVert \left( \sum_{v_{2}} (-1)^{\delta_{v_{1}=v_{2}}} B_{v_{2}} \right)  \right\rVert^{2}_{\rho} }.
    \end{align*}
    
    Focusing on the last factor in the last term,
    
    \begin{align*}
        \sum_{v_{1}} \norm{ \brac{ \sum_{v_{2}} (-1)^{\delta_{v_{1}=v_{2}}} B_{v_{2}} }  }^{2}_{\rho} &= \sum_{v_{1}} \sum_{v_{2}, v_{2}^{\prime}} (-1)^{\delta_{v_{1}=v_{2}}+\delta_{v_{1}=v_{2}^{\prime}}} \Tr{ B_{v_{2}}, B_{v_{2}^{\prime}} \rho} \\
            &= n^{2} + \sum_{v_{1}} \sum_{v_{2}\neq v_{2}^{\prime}} (-1)^{\delta_{v_{1}=v_{2}}+\delta_{v_{1}=v_{2}^{\prime}}} \Tr{ B_{v_{2}}, B_{v_{2}^{\prime}} \rho} \\
            &= n^{2} + \sum_{v_{2}\neq v_{2}^{\prime}} \left( \sum_{v_{1}} (-1)^{\delta_{v_{1}=v_{2}}+\delta_{v_{1}=v_{2}^{\prime}}} \right) \Tr{ B_{v_{2}}, B_{v_{2}^{\prime}} \rho} \\
            &= n^{2} + \sum_{v_{2} \neq v_{2}^{\prime}} (n-4) \Tr{ B_{v_{2}}, B_{v_{2}^{\prime}} \rho}\\
            &= n^{2} - (n-4)n + (n-4)\left\lVert \sum_{v_{2}} B_{v_{2}} \right\rVert^{2}_{\rho}.
    \end{align*}
    
    Then, the above becomes
    \begin{align*} 
        \frac{1}{2n^{2}} \left\lVert n^{2} - \sum_{v_{1}} A_{v_{1}} \left( \sum_{v_{2}} (-1)^{\delta_{v_{1}=v_{2}}} B_{v_{2}} \right)  \right\rVert_{\rho} &= \frac{1}{2} + \frac{1}{2n^{2}}\sqrt{n}\sqrt{ n^{2} - (n-4)n + (n-4)\left\lVert \sum_{v_{2}} B_{v_{2}} \right\rVert^{2} } \\
            &= \frac{1}{2} + \frac{1}{2n^{2}}\sqrt{n} \sqrt{ 4n + (n-4)\left\lVert \sum_{v_{2}} B_{v_{2}} \right\rVert^{2} }
    \end{align*}
    
    For $n=3$, $(n-4) = -1$, the above quantity is upperbounded by
    
    $$ \frac{1}{2} + \frac{1}{2n^{2}}\sqrt{n} \sqrt{ 4n } = \frac{5}{6}. $$
    
    For $n \geq 4$, we need to upperbound $\left\lVert \sum_{v{2}} B_{v_{2}} \right\rVert^{2}$ by $n^{2}$. This gives us
    
    $$ \frac{1}{2} + \frac{\sqrt{n^{2} - 4n +4}}{2n} = \frac{1}{2} + \frac{|n-2|}{2n} $$
    
    which is $ 1 - \frac{1}{n} $ for $n \geq 4$. 
\end{proof}

The synchronous quantum strategy saturates the $\frac{5}{6}$ bound for $K_{3,2}$.

For $BPM_{K_{n,2}}$ for $n \geq 4$ we show that we have classical strategies which saturate the quantum upperbound. 

\begin{lem}
    $\omega(BPM_{K_{n,2}}) \geq 1 - \frac{1}{n}$ for all $n \geq 4$.
\end{lem}
\begin{proof}
    Consider the ``trivial'' classical strategy where Alice always outputs the first right vertex and Bob always outputs the second right vertex. We note that this strategy only fails when Alice and Bob receive the same vertex. This constitutes $n$ of the $n^{2}$ vertex pairs that the players could be asked which establishes the result.
\end{proof}

We now finally give a classical upperbound for $BPM(K_{3,2})$ which completes the study of the quantum and classical values for $K_{n,2}$ bipartite perfect matching games. 

\begin{lem}
    $\omega(BPM_{K_{3,2}}) = \frac{7}{9}$.
\end{lem}
\begin{proof}
    The optimal classical value can be achieved via a deterministic strategy. Let $a_{i}, b_{j} \in \cbrac{\pm 1}$ be the expected value for the deterministic strategy with $+1$ be the weighting for the first right vertex and $-1$ be the weighting for the second right vertex. 
    
    Without loss of generality, $a_{1} = a_{2}$. Then, 
    \begin{align*}
        \frac{1}{18} \brac{9 - a_{1}\brac{- b_{1} + b_{2} + b_{3}} - a_{2}\brac{b_{1} - b_{2} + b_{3}} - a_{3}\brac{b_{1} + b_{2} - b_{3}} } &= \frac{1}{18} \brac{9 - 2a_{1}b_{3} - a_{3}\brac{b_{1} + b_{2} - b_{3}} } \\
            &\leq \frac{14}{18} = \frac{7}{9}.
    \end{align*}
    We see that this is optimal as setting $a_{1}=a_{2} = -1$, $a_{3}=1$, $b_{1}=b_{2} = -1$, and $b_{3} = 1$ yields the value $\frac{7}{9}$.
\end{proof}

This shows that we only have quantum advantage for $n=3$ for the graph $K_{3,2}$, which is also the only case where the optimal quantum value is achieved by playing synchronously. We summarize all the bounds in the theorem below.

\begin{thm}
    For the $BPM_{K_{n,2}}$ games, we have the following:
    \begin{enumerate}
        \item $\omega(BPM_{K_{3,2}}) = \frac{7}{9} < \frac{5}{6} = \omega^{\ast}(BPM_{K_{3,2}})$,
        \item $\omega(BPM_{K_{n,2}}) =\omega^{\ast}(BPM_{K_{n,2}}) = 1 - \frac{1}{n}$ for all $n \geq 4$,
        \item $\omega^{\ast,s}(BPM_{K_{n,2}}) = \frac{1}{2} + \frac{1}{n}$ for all $n \geq 2$.
    \end{enumerate}
\end{thm}

Additionally, we have a sum-of-squares decomposition for $BPM_{K_{3,2}}$ which gives an alternative proof of $\omega^{\ast}(BPM_{K_{3,2}}) = \frac{5}{6}$. 
\begin{align*} 
6I - &( A_{1}(B_{1}-B_{2}-B_{3}) + A_{2}(-B_{1} + B_{2} - B_{3}) + A_{3}(-B_{1} - B_{2} + B_{3} ) ) \\
        &\qquad= \frac{1}{2} \brac{ A_{1} - A_{3} - B_{1} + B_{3} }^{2} + \frac{1}{2} \brac{ A_{1} - A_{2} - B_{1} + B_{2} }^{2}  \\
        &\qquad\qquad+ \frac{1}{4} \brac{ A_{1} + A_{2} + A_{3} + B_{1} + B_{2} + B_{3} }^{2} + \frac{1}{12}\brac{ A_{1} + A_{2} + A_{3} - B_{1} - B_{2} - B_{3} }^{2}
\end{align*}

Furthermore, any optimal strategy for this game must satisfy the identities $\sum_i A_i = \sum_i B_i = 0$. This can be derived from the sum-of-square proof of optimality by adding the last two terms to get $$(A_{1} + A_{2} + A_{3} + B_{1} + B_{2} + B_{3}) + (A_{1} + A_{2} + A_{3} - B_{1} - B_{2} - B_{3}) = 2\sum_i A_i = 0$$ and similarly, $$(A_{1} + A_{2} + A_{3} + B_{1} + B_{2} + B_{3}) - (A_{1} + A_{2} + A_{3} - B_{1} - B_{2} - B_{3}) = 2\sum_i B_i = 0.$$

%% file: sections/PerfectMatching.tex
\section{Perfect matching}\label{NS}

In this section, we take the same approach to define the notions of quantum and nonsignaling perfect matching for general graphs.

\subsection{Quantum perfect matching}
Unlike the bipartite and fractional games, the perfect matching game exhibits quantum pseudo-telepathy for many graphs. Therefore quantum perfect matching defines a distinct property to perfect matching.

First we give a simple characterization of quantum strategies for the perfect matching game.

\begin{lem} \label{lem:bipartite-strategies}
Given a graph $G=(V,E)$ any perfect synchronous quantum strategy for $PM_G$ given by projectors ${\Pi_x^y}$ for $x,y \in V$ will have the following properties:
\begin{enumerate}
    \item $\Pi_{(x,y)} := \Pi_x^y = \Pi_y^x$ for every $x,y \in V$.
    \item $\Pi_{e} = 0$ if  $e \not \in E$.
    \item $\sum_{e \in E(x)} \Pi_e = I$ for every $x \in V$.
    \item $\Pi_e \Pi_h = \delta_{e=h}\Pi_{e}$ for every $e,h \in E$ where $e\cap h \neq \emptyset$.
\end{enumerate}
\end{lem}
\begin{proof}
Rule games imply (1,2,4), (3) because its a strategy.
\end{proof}

We will see that a graph $G$ having a quantum perfect matching is equivalent to its line graph $L(G)$ having a projective packing of value $|V(G)|/2$, which is equivalent to $2L(G)$ (the disjoint union of two copies of $L(G)$) having quantum independence number $|V(G)|$. First we must define projective packing number and quantum independence number.

\begin{defn}[\cite{MRVdeciding, robersonthesis}]
A \emph{$d$-dimensional projective packing} of a graph $G$ is an assignment $x \to \Pi_x \in \mathbb{C}^{d \times d}$ of projections to the vertices of $V(G)$ such that $x \sim y$ implies that $\Pi_x\Pi_y = 0$. The \emph{value} of such a projective packing is
\[\frac{1}{d}\sum_{x \in V(G)} \mathrm{rk}(\Pi_x) = \frac{1}{d}\sum_{x \in V(G)} \mathrm{Tr}(\Pi_x) = \frac{1}{d}\mathrm{Tr}\left(\sum_{x \in V(G)} \Pi_x\right).\]
Additionally, the \emph{projective packing number} of a graph $G$, denoted $\alpha_p(G)$ is the supremum of the values of projective packings of $G$.
\end{defn}

The quantum independence number of a graph $G$ is defined as the maximum $k$ such that there is a perfect quantum strategy for the $k$-independent set game on $G$. However, this is equivalent to the following definition which will be of more direct use for us.

\begin{defn}[\cite{quantumHom, robersonthesis}]
    Let $G$ be a graph. The \emph{quantum independence number} of $G$, denoted $\alpha_q(G)$, is the maximum $k \in \mathbb{N}$ such that there exists finite dimensional projections $\Pi_{i,x}$ for all $i \in [k]$ and $x \in V(G)$ satisfying the following:
    \begin{enumerate}
        \item $\sum_{x \in V(G)} \Pi_{i,x} = I$ for all $i \in [k]$.
        \item $\Pi_{i,x}\Pi_{j,y} = 0$ if $i \ne j$ and $x=y$ or $x \sim y$.
    \end{enumerate}
    Note that the first condition implies that $\Pi_{i,x}\Pi_{i,y} = 0$ if $x \ne y$.
\end{defn}

It is known that if projections $\Pi_{i,x}$ for $i \in [k]$ and $x \in V(G)$ give a quantum $k$-independent set of $G$, then the projections $\Pi_x := \sum_{i \in [k]} \Pi_{i,x}$ are a projective packing of $G$ of value $k$ and thus $G$ always has a projective packing of value equal to its quantum independence number and thus $\alpha_q(G) \le \alpha_p(G)$.

We now show that a quantum perfect matching of $G$ is equivalent to a projective packing of $L(G)$ of value $|V(G)|/2$. We remark that this is the maximum possible value of a projective packing of $L(G)$.

\begin{thm} \label{mainthm}
    Let $G$ be a graph. Then the following are equivalent:
    \begin{enumerate}
        \item $G$ has a quantum perfect matching.
        \item $L(G)$ has a projective packing of value $|V(G)|/2$.
        \item $\alpha_q(2L(G)) = |V(G)|$.
    \end{enumerate}
\end{thm}
\begin{proof}
    We will show that $(1) \Leftrightarrow (2)$ ( $(2) \Leftrightarrow (3)$ follows from the above discussion). We first prove that if $G$ has a quantum perfect matching, then $L(G)$ has a projective packing of value $|V(G)|/2$. Since $G$ has a quantum perfect matching, by Lemma~\ref{lem:bipartite-strategies} there is an assignment $e \to \Pi_e \in \mathbb{C}^{d \times d}$ (for some $d$) of projections to the edges of $G$ such that $\Pi_e\Pi_f = 0$ if $e$ and $f$ are incident and $\sum_{e \in E(X)}\Pi_e = I$ for all $x \in V(G)$. Since incidence of edges is adjacency in the line graph, we have that this is in fact a projective packing of $L(G)$. To compute its value, note that 
    \[2 \frac{1}{d} \mathrm{Tr}\left(\sum_{e \in E(G)} \Pi_e \right) = \frac{1}{d} \mathrm{Tr}\left(\sum_{x \in V(G)} \sum_{e \in E(x)} \Pi_e \right) = \frac{1}{d}\mathrm{Tr}(|V(G)|I) = |V(G)|.\]
    Therefore the projective packing of $L(G)$ has value $|V(G)|/2$.

    Now suppose that $L(G)$ has a projective packing $e \to \Pi_e \in \mathbb{C}^{d \times d}$ of value $|V(G)|/2$. Note that these projections satisfy the orthogonality requirements of Lemma~\ref{lem:bipartite-strategies} and thus to prove that $G$ has a quantum perfect matching it is only left to show that $\sum_{e \in E(x)} \Pi_e = I$ for all $x \in V(G)$. For any $x \in V(G)$, we have that $\Pi_{e}\Pi_f = 0$ for any two distinct $e,f \in E(x)$. Therefore, $\sum_{e \in E(x)}$ is a projection and thus
    \[\mathrm{Tr}\left(\sum_{e \in E(x)}\right) \le \mathrm{Tr}(I) = d,\]
    with equality if and only if the sum is equal to the identity. Therefore,
    \[\frac{1}{d}\mathrm{Tr}\left(\sum_{e \in V(L(G))} \Pi_e\right) = \frac{1}{2d}\mathrm{Tr}\left(\sum_{x \in V(G)} \sum_{e \in E(x)} \Pi_e\right) \le |V(G)|/2,\]
    with equality if and only if $\sum_{e \in E(x)} \Pi_e = I$ for all $x \in V(G)$. But we do have equality by assumption, and therefore we have proven that $G$ has a quantum perfect matching. We remark that the above also shows that $|V(G)|/2$ is always an upper bound on the value of a projective packing of $L(G)$, and so $G$ has a quantum perfect matching precisely when this maximum possible value is attainable.

\end{proof}

\begin{lem}
    $K_{5}$ has no quantum pseudotelepathy.
\end{lem}
\begin{proof}
    We shall show that there cannot be projectors satisfying the above properties for $K_{5}$. We begin by showing that if all the above properties are satisfied then all of the projectors must be orthogonal. Indeed, consider edges $(x,y)$ and $(z,w)$ which do not share a vertex. Then, there is a unique vertex $a \neq x,y,z,w$ and so
    \[ \Pi_{(x,y)} = \Pi_{(x,y)}\brac{ \sum_{i \neq a} \Pi_{(i,a)} } = \Pi_{(x,y)}\brac{ \Pi_{(z,a)} + \Pi_{(w,a)} }. \]
    Therefore, 
    \[ \Pi_{(x,y)}\Pi_{(z,w)} = \Pi_{(x,y)}\brac{ \Pi_{(z,a)} + \Pi_{(w,a)} }\Pi_{(z,w)} = 0.\] 
    Now, for any edge $(x,y)$, 
    \[ I = \brac{ \sum_{a \neq y} \Pi_{(a,y)} }\brac{ \sum_{b \neq x} \Pi_{(x,b)} } = \Pi_{(x,y)}^{2} = \Pi_{(x,y)} \]
    which is impossible.
\end{proof}

\begin{lem}
    $\omega^{*}(PM_{K_7}) = 1$.
\end{lem}
\begin{proof}
    A projective packing of $L(K_7)$ is provided by the 7-context Kochen-Specker sets in Section 2 of \cite{KS}.
\end{proof}

\begin{thm}
    $\omega^{*}(PM_{K_n}) = 1$ if and only if $n \neq 1,3,5$.
\end{thm}
\begin{proof}
    Since $K_{5}$ does not have a quantum perfect matching, neither does $K_{3}$. $K_{n}$ has a quantum perfect matching for odd $n \geq 7$ as it can be broken into $K_7$ and an even number of additional verticies that can be classically matched.
\end{proof}

\subsection{Nonsignaling perfect matching}
The cycle graphs $C_n$  for odd $n \geq 5$ are an infinite family of graphs with nonsignaling perfect matching and no quantum or classical perfect matching.

\begin{thm}
For odd $n \geq 5$, $\omega^{ns}(PM_{C_n}) = 1$. \label{cycle_ns}
\end{thm}
\begin{proof}
The following defines a simple perfect nonsignaling strategy $p$. Here $\to$ and $\leftarrow$ represent the two edges at any vertex of the cycle graph where some orientation of the edges is fixed.
\begin{enumerate}
    \item $p(\to,\to,x,x) = p(\leftarrow,\leftarrow,x,x) =  \frac{1}{2}$ for every vertex $x$.
    \item $p(\to,\leftarrow,x,y) = p(\leftarrow,\to,x,y) =  \frac{1}{2}$ for any two adjacent vertices $x \sim y$.
    \item $p(\to,\to,x,y) = p(\leftarrow,\leftarrow,x,y) = p(\to,\leftarrow,x,y) = p(\leftarrow,\to,x,y)  =  \frac{1}{4}$ for any two disconnected vertices $x \nsim y$.
\end{enumerate}
\end{proof}

\begin{defn}
    Given graph $G$, we say that a fractional perfect matching $f : E(G) \to \R$ \emph{avoids triangles} if $\sum_{e \in t} f(e) \leq 1$ for all triangles $t$ in $G$.
\end{defn}


\begin{thm} \label{thm:ns_characterization}
    For a graph $G$, $\omega^{ns}(PM_{G}) = 1$ if and only if $G$ has a fractional perfect matching avoiding triangles.
\end{thm} 
\begin{proof}
    In this proof, it will be convenient to consider vertices instead of edges for the answers where responding to the question $x$ with vertex $a$ specifies the edge $(a, x)$.

    Suppose $\omega^{ns}(PM_{G}) = 1$ and let $p$ be a bisynchronous nonsignaling correlation witnessing this. We will show that $f(a,x) := p(a|x)$, the marginal of $p$ defines a fractional perfect matching avoiding triangles. Firstly, since $p$ is nonsignaling, for any $a, x \in V(G)$ and any $y \in V(G)$,
    \[ p(a|x) = \sum_{b \in N(y)} p(a, b | x, y). \]
    Substituting $y = x$, we have
    \[ p(a|x) = \sum_{b \in N(x)} p(a, b| x, x) = p(a, a|x, x) \]
    and substituting $y = a$, we have
    \[ p(a|x) = \sum_{b \in N(a)} p(a, b| x, a) = p(a, x|x,a). \]
    This shows that
    \[ p(a|x) = p(a, a|x,x) = p(a, x|x,a) = p(x,x|a,a) =  p(x|a). \]
    
    Now, fix $x \in V(G)$, then 
    \[ \sum_{(a,x) \in E(x)} f(a,x) = \sum_{(a,x) \in E(x)} p(a|x) = \sum_{(a, x) \in E(x)} p(a, a|x,x) = 1. \]
    Hence, $f$ is a fractional perfect matching. Now, fix a triangle $\cbrac{v, u, w} \subset V(G)$. Then,
    \begin{align*}
        1 &= \sum_{a \in N(v), b \in N(u)} p(a, b|v,u) \\
            &\geq p(u, v | v, u) + \sum_{a \in N(v)} p(a, w | v, u) + \sum_{b \in N(u)} p(w, b | v, u) \\
            &= p(u|v) + p(w|u) + p(w|v) \\
            &= f(u, v) + f(w, u) + f(w, v),
    \end{align*}
    where we used the fact that marginals are well-defined from the nonsignaling condition from line 2 to 3. Hence, $f$ is bounded by 1 on triangles. This completes the forward direction. 

    Now, for the backwards direction note that the existence of a fractional perfect matching can be formulated as feasibility of a linear program with only integer coefficients. Moreover, the triangle-avoiding condition can also be encoded as linear constraints with integer coefficients. Therefore, if $G$ has a fractional perfect matching that avoids triangles, then it has one that is rational-valued. Let $f:E(G) \to \mathbb{Q}$ be a rational-valued fractional perfect matching of $G$. It follows that there is some value $r \in \mathbb{N}$ such that $h(e) := rf(e) \in \mathbb{Z}^{\ge 0}$ for all $e \in E(G)$. Note that this means that $\sum_{y \in N(x)} h(xy) = r$ for all $x \in V(G)$, and that $\sum_{e \in t} h(e) \le r$ for all triangles $t$ in $G$. We will show how to use $f$ to construct a perfect nonsignaling correlation $p$ for the perfect matching game for $G$ whose marginals $p(y|x)$ are equal to $f(xy)$ for all $xy \in E(G)$.

    First, define
    \[p(y,y'|x,x) = \begin{cases} f(xy) & \text{if } y = y' \text{ and } xy \in E(G)\\ 0 & \text{otherwise}\end{cases}\]
    
    Now let $x,x' \in V(G)$ with $x \ne x'$. If $xx' \in E(G)$, let $k = h(xx')$, and otherwise set $k = 0$. Note that for any $y \in N(x) \cap N(x')$, we have that $h(xy) + h(x'y) \le r-k$. Define a bipartite graph $H$ with parts $A = \{(y, i, 0) : y \in N(x)\setminus \{x'\}, \ i = 1, \ldots, h(xy)\}$ and $B = \{(y, i, 1) : y \in N(x')\setminus \{x\}, \ i = 1, \ldots, h(x'y)\}$, where $(y,i,0) \sim (y',j,1)$ if $y \ne y'$. Note that $|A| = |B| = r-k$. We will show that $H$ has a perfect matching via Hall's theorem. Consider a subset $S \subseteq A$, and define $T = \{y \in N(x) \setminus \{x'\} : \exists \ i \in [h(xy)] \text{ s.t. } (y,i,0) \in S \text\}$. Note that if $|T| > 1$, then $N(S) = B$ and thus Hall's condition holds for $S$. Of course $|T| = 0$ if and only if $S = \varnothing$, so we may assume that $|T| = 1$ and thus there is some $\hat{y} \in N(x) \setminus \{x'\}$ such that $S \subseteq \{(\hat{y},i,0) : i \in [h(x\hat{y})]\}$. Thus $|S| \le h(x\hat{y})$. Moreover, every vertex of $S$ has the same neighborhood, which is $B \setminus \{(\hat{y},j,1) : j \in [h(x'\hat{y}]\}$. Therefore,
    \[|N(S)| = |B| - h(x'\hat{y}) = r-k - h(x'\hat{y}).\]
    Recalling that $h(x\hat{y}) + h(x'\hat{y}) \le r-k$, we see that
    \[|S| \le h(x\hat{y}) \le r-k - h(x'\hat{y}) = |N(S)|.\]
    Thus Hall's condition holds for arbitrary $S \subseteq A$ and therefore $H$ has a perfect matching.
    
    Now let $M$ be a perfect matching of $H$. For each $y \in N(x) \setminus \{x'\}$ and $y' \in N(x') \setminus \{x\}$, define $g(y,y')$ to be the number of edges in $M$ of the form $(y,i,0)(y',j,1)$ for $i \in [h(xy)]$ and $j \in [h(x'y')]$. Then define
    \[p(y,y'|x,x') = \begin{cases} f(xx') & \text{if } y=x', \ y' = x, \ \& \ xx' \in E(G)\\ g(y,y')/r & \text{if } y \in N(x) \setminus \{x'\} \ \& \ y' \in N(x') \setminus \{x\}\\ 0 & \text{otherwise.}\end{cases}\]
    It is now straightforward to check that $p$ is a valid correlation that wins the perfect matching game for $G$ with probability 1, and that the marginals $p(y|x)$ are well-defined and equal to $f(xy)$ if $xy \in E(G)$ and equal to 0 otherwise. 
\end{proof}

\subsection{Undecidability of quantum perfect matching for hypergraphs} \label{und-hyper}

As we saw in \cref{hypergraph_def}, the perfect matching game can be also played on hypergraphs. Unlike with graphs for which it remains open whether quantum perfect matching is decidable, for hypergraphs we get undecidability.

\begin{thm}
    It is undecidable to decide if a hypergraph has a quantum perfect matching. 
\end{thm}
\begin{proof}
    From \cref{mainthm}, we get that deciding quantum perfect matching for a hypergraph $H$ is equivalent to deciding $\alpha_q(2L(H))$. A classical result in graph theory tell us that the set of line graphs of hypergraphs is equivalent to that of general graphs \cite{hyper}. That is given a graph $G$ there exists a hypergraph $H$ such that $L(H) = G$. Therefore, deciding quantum perfect matching for hypergraphs is equivalent to deciding if a graph has a quantum independence number of $|V|$ which is undecidable \cite{harris2023universalitygraphhomomorphismgames}.
\end{proof}